\documentclass[11pt,twocolumn]{article}
\usepackage{amssymb,epsfig}
\usepackage{color}
\def\comment#1{{}}

\newlength{\cvindent}
\newlength{\cvhang}

\newlength{\refindent}
\begin{document}

\vspace{-0.5in}

%\begin{center} 
%\textbf{\Large 
\title{The Status and future of ground-based TeV gamma-ray astronomy\\
Reports of Individual Working Groups}
\author{M. Pohl, Iowa State University,\\
A. Abdo, Michigan State University/NRL,\\
A. Atoyan, Montreal University,\\
M. Baring, Rice University,\\
J. Beacom, Ohio State University,\\ 
R. Blandford, KIPAC/SLAC,\\
Y. Butt, The Harvard Smithsonian Center for Astrophysics,\\
A. Bykov, Ioffe Physico-Technical Institute, St. Petersburg,\\
D. Ellison, North Carolina State University,\\
S. Funk, KIPAC/SLAC,\\
F. Halzen, University of Wisconsin, Madison,\\
E. Hays, NASA/GSFC,\\
B. Humensky, The University of Chicago,\\
T. Jones, University of Minnesota,\\
P. Kaaret, The University of Iowa,\\
D. Kieda, The University of Utah,\\
S. LeBohec, The University of Utah,\\
P. Meszaros, Pennsylvania State University,\\
I. Moskalenko, SLAC,\\
P. Slane, Harvard-Smithsonian Center for Astrophysics,\\
A. Strong, Max-Planck-Institut f\"ur extraterrestrische Physik,\\
S. Wakely, The University of Chicago}
\date{}
\maketitle
\section{Galactic diffuse emission, supernova remnants, and the origin of cosmic rays}
\label{GDE-subsec}
\subsection{Why are they important?}
The origin of Galactic cosmic rays and the mechanisms of their acceleration are
among the most challenging problems in astroparticle physics and also
among the oldest. Cosmic rays are energetically important in our
understanding of the interstellar medium (ISM) because they contain at
least as much energy as the other phases of the ISM.  They also provide, along with
interstellar dust, the only sample of ordinary matter from outside the heliosphere. Yet, the origin
of cosmic rays in the Galaxy remains uncertain more than 90 years after their discovery by
Victor Hess in 1912 (for a recent review, see \cite{hillas05}). 
Improving our knowledge of the interaction between highly
energetic particles and the other elements of the ISM could help
understand other systems, such as active galactic
nuclei (AGN) that produce
strong outflows with highly energetic particles.

High-energy gamma rays are a unique probe of cosmic rays. Observations in the TeV band 
are a sensitive probe of the highest energy physical processes occurring in a 
variety of astronomical objects, and they allow us to measure the properties of energetic 
particles anywhere in the Universe, such as their number, composition, and 
spectrum. From such measurements we know already
that our Galaxy contains astrophysical systems capable of accelerating particles to
energies beyond the reach of any accelerator built by humans. What drives these
accelerators is a major question in physics and understanding these accelerators has broad 
implications, but more sensitive gamma-ray detectors are needed to address these questions.
Among the many types of Galactic
gamma-ray sources, observations of high-energy emission from shell-type 
supernova remnants (SNR) are particularly beneficial because:

\begin{itemize}
\item The acceleration of relativistic charged particles is one of the main unsolved,
yet fundamental, problems in modern astrophysics. Only in the case of SNRs do we have an 
opportunity to perform
spatially resolved studies in systems with known geometry, and the plasma physics 
deduced from these observations will help us to understand other systems where rapid particle 
acceleration is believed to occur and where observations as detailed as those of SNRs are 
not possible.

\item The acceleration of particles relies on interactions between
energetic particles and magnetic turbulence, so the question of cosmic-ray acceleration is, in fact,
one of the generation, interaction, and damping of turbulence in a non-equilibrium plasma. 
The physics of the coupled system of turbulence, energetic particles, and colliding plasma flows 
can be ideally studied in young SNRs, for which observations in X-rays 
\cite{Koyama95} and TeV-scale gamma rays \cite{aha04} indicate a
very efficient particle acceleration to at least 100 TeV and the existence of a
turbulent magnetic field that is much stronger than a typical shock-compressed
interstellar magnetic field. The amplification of magnetic fields by streaming
energetic particles is of particular
interest because it may play an important role in the generation of cosmological magnetic fields. 
\item SNR are the most likely candidate for the sources of cosmic rays, either
as isolated systems or acting collectively in groups in so-called superbubbles, although to date
we do not have conclusive evidence that they produce cosmic-ray ions in addition to
electrons. An understanding of
particle acceleration in SNR may solve the century-old question of the
origin of cosmic rays.
\item SNR are a major source of heat and turbulence in the interstellar medium of galaxies, 
and thus have an impact on the evolution of the galactic ecosystems. 
In particular, when new insights are 
extended to shocks from other sources, e.g. the winds of massive stars, they will help in
advancing our understanding of the energy balance and evolution of the interstellar medium in
galaxies. 
\item The evolution and interaction of turbulence and cosmic rays determines how the cosmic rays
will eventually be released by the SNR, which has an impact on the amplitude and frequency
of variations of the cosmic-ray flux near Earth and at other locations in the Galaxy
\cite{pe98}.  
\end{itemize}

The study of diffuse Galactic gamma-ray emission is important for a number of reasons.

\begin{itemize}
\item It provides direct information on the cosmic-ray spectrum in various 
locations in the Galaxy, which is needed to understand the origin of cosmic rays 
near and beyond the knee.

\item It must be understood to properly analyze extended gamma-ray sources,
in particular in terms of possible spatial variations of its
spectrum resulting from non-stationary cosmic ray transport. 

\item It will enable us to analyze the gamma-ray spectra of supernova remnants
self-consistently in the 
light of their function as possible sources of Galactic cosmic rays.

\item It allows us to derive self-consistent limits on the amount of
dark matter in the Galaxy by determining both the cosmic ray propagation and 
radiation properties
and the gamma-ray emissivities of dark matter for a variety of spatial distributions.

\end{itemize}

\subsection{What do we know already?}

\medskip
\subsubsection{Supernova remnants}

Cosmic rays consist of both electrons and hadrons. However, the hadrons
dominate the energy budget, and the acceleration of hadrons is the key
issue in understanding the origin of cosmic rays. The energy density in
local cosmic rays, when extrapolated to the whole Galaxy, implies the
existence of powerful accelerators in the Galaxy.  Supernova remnants
(SNRs) have long been thought to be those accelerators
\cite{ginzburg64}, but there is no definitive proof that hadrons are
accelerated in SNRs. The classical argument that shocks in shell-type
SNRs accelerate cosmic rays is that supernova explosions are
one of the few Galactic phenomena capable of satisfying the energy
budget of cosmic-ray nuclei, although even supernovae must have a high
efficiency ($\sim$10\%-20\%) for converting the kinetic energy of the
SNR explosions to particles \cite{drury89}. However, these arguments are indirect. 
Other source
classes may exist that have not been considered to date, and one may ask
what role is played by the many sources seen in the TeV-band that do not have 
an obvious counterpart in other wavebands \cite{hess-survey}. In any case,
observations of TeV photons from SNRs offer the most promising direct way to confirm 
whether or not SNRs are, in fact, the main sources of CR ions below
$10^{15}$~eV. 

\begin{figure*}[tb]
\centerline{\psfig{file=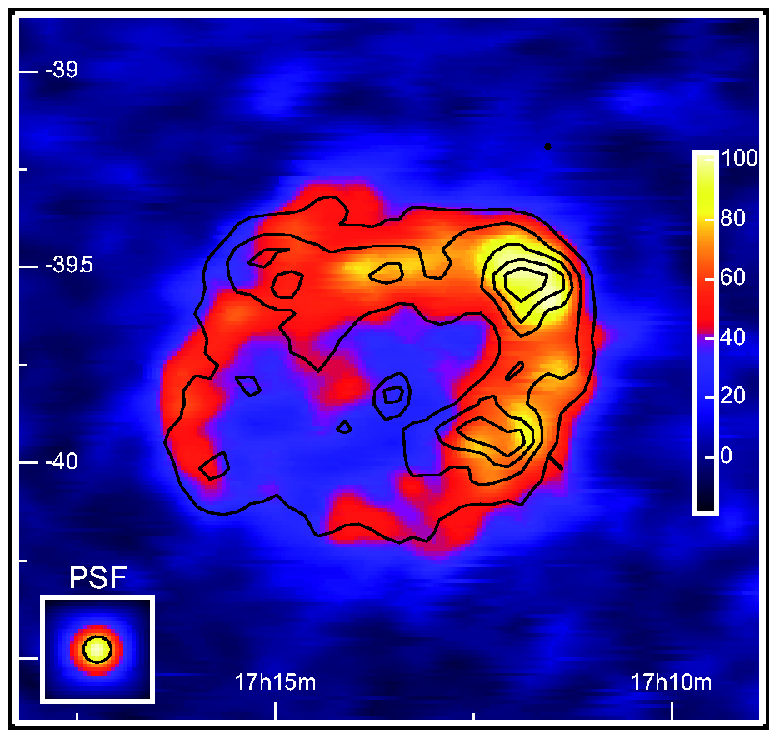,width=2.8in,clip=} \hspace{0.2in}
\psfig{file=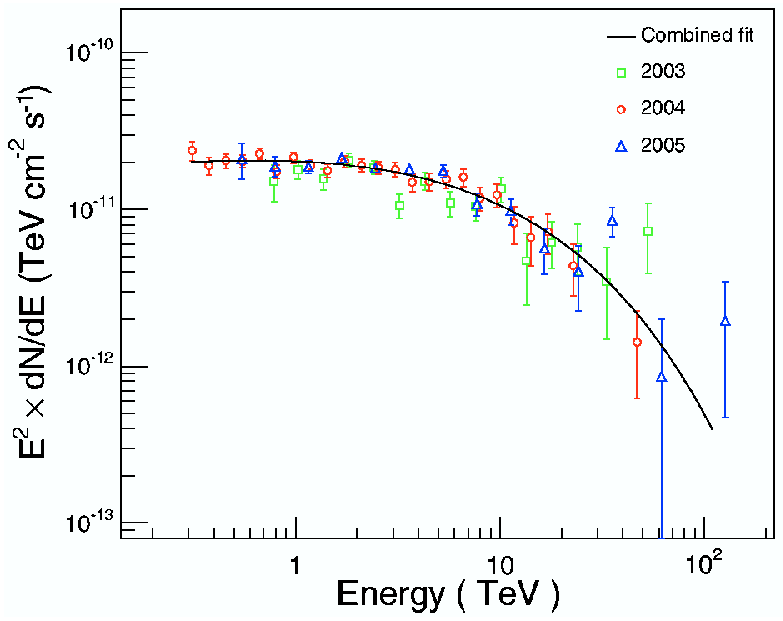,width=3.3in,clip=}} 
\caption{The left panel shows an image of the acceptance-corrected
gamma-ray excess rate in the TeV band as observed with H.E.S.S. from the SNR
RX~J1713-3946 \cite{hess-rx}.
The insert labeled PSF indicates how a point source would appear in this image. Overlaid are
black contour lines that indicate the X-ray intensity at 1-3 keV. Note the similarity between the
X-ray and TeV-band images. The right panel shows the TeV-band spectrum for the entire remnant
broken down for three different observing seasons.} \label{ic443} 
\end{figure*}

Even though very early measurements showed that the fluxes of TeV emission from SNRs
are lower than originally predicted if SNRs really do accelerate the
bulk of Galactic cosmic rays \cite{buckley98}, later observations with H.E.S.S.
established shell-type SNRs such as RX~J1713-3946 \cite{hess-rx} and
RX~J0852.0-4622 \cite{hess-vj} as TeV-band gamma-ray sources. The maturity of high-energy
gamma-ray astrophysics is best illustrated by the ability of current atmospheric Cherenkov detectors
such as H.E.S.S., MAGIC, and VERITAS
to resolve sources and to map the brightness distribution in TeV-band gamma rays. 
Figure~\ref{ic443} shows such a gamma-ray map and the TeV-band spectrum of RX~J1713-3946.
The interpretation of these TeV observations is complicated because two competing 
radiation processes, pion-decay photons from ion-ion interactions and 
Inverse-Compton
(IC) emission from TeV electrons scattering off the cosmic microwave background and 
the ambient galactic radiation, 
can produce similar fluxes in the GeV-TeV energy range. 
In the hadronic scenario neutrinos
would be produced through the decay of charged pions. If even a few
neutrinos are detected from a source at high enough energies, where the atmospheric neutrino background is minimal, then this alone
could decisively indicate the hadronic mechanism \cite{kis+bea}. 

\begin{figure*}[tb] \centerline{\psfig{file=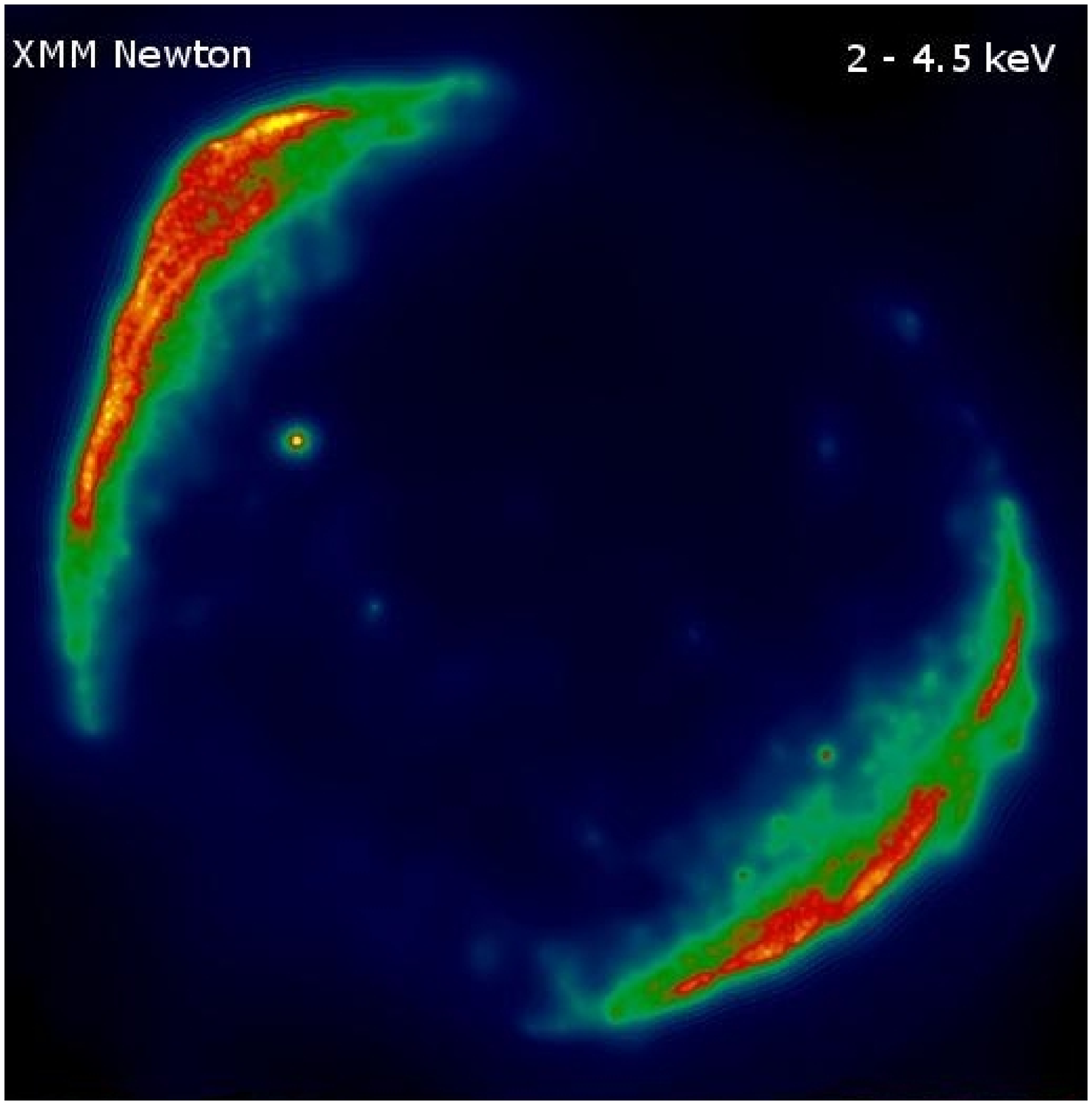,width=3.0in,clip=} ~
\psfig{file=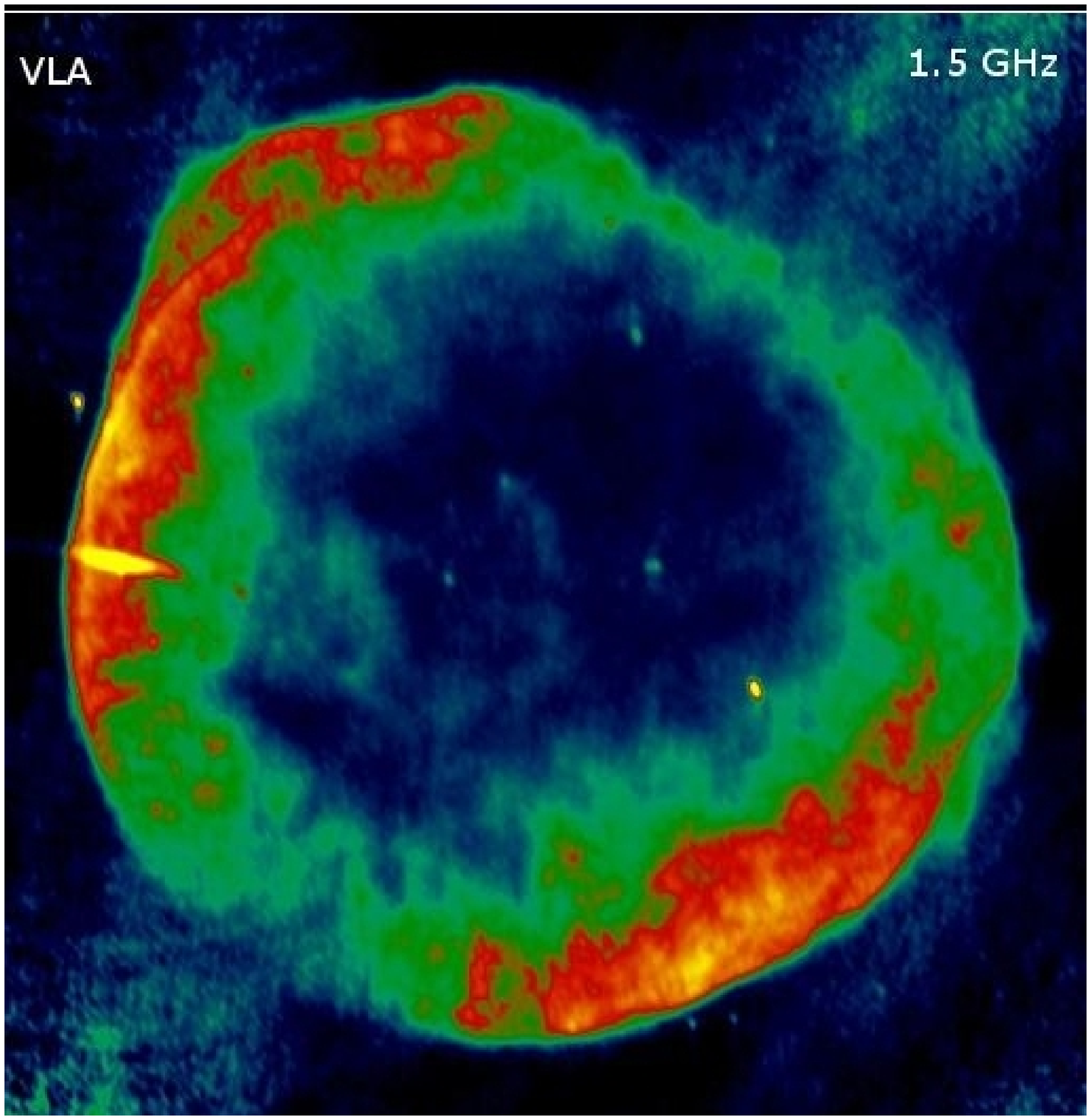,width=3.in,clip=}} \caption{X-ray and radio
images of SN 1006 \cite{rothenflug04}. Hard X-rays (left) are mainly
produced by very high-energy electrons ($\sim 100$~TeV) emitting
synchrotron radiation.  Radio emission (right) is produced by electrons
with energies in the GeV range emitting synchrotron radiation.  Imaging
TeV observations will enable us to map the inverse-Compton emission
from high-energy electrons and make a measurement of the magnetic field
strength in the vicinity of the shock.  Such mapping is also essential
for distinguishing TeV photons produced by electronic versus hadronic
cosmic rays.  The angular size of the image is 35~arcmin.  (Image
courtesy of CEA/DSM/DAPNIA/SAp and ESA.)} \label{sn1006} \end{figure*}

SNRs do accelerate {\it electrons}. As has long been known from radio observations, 
GeV-scale electrons are accelerated in SNRs, and now compelling evidence
for acceleration of electrons at the forward shocks of SNRs comes from observations of
non-thermal X-rays from several shell-type SNRs.  The X-ray
emission is synchrotron radiation from electrons accelerated to TeV
energies. In the case of SN 1006, the electrons must have energies of
at least 100~TeV \cite{Koyama95}, see Fig.~\ref{sn1006}.  These
electrons must be accelerated in situ because such energetic electrons
cannot travel far from their origin before they are attenuated by energy losses
due to synchrotron radiation.  The same electrons should produce TeV emission via
inverse-Compton scattering. The intensity and spectrum of the emission
are determined by the electron density, maximum electron energy, and
local magnetic field. Combining radio, X-ray, and TeV data can provide a
measurement of the magnetic field strength in the vicinity of the
shock.  This important parameter is not provided by X-ray
spectroscopy alone, because the photon cut-off energy is insensitive to the
magnetic field strength if it is generated by the competition of
strong synchrotron cooling and gyroresonant acceleration of
electrons. 

An important clue to the nature of the parent particles comes
from correlation studies with X-rays in the 2--10~keV band. For the
two prominent SNRs, RX\,J1713.7-3946 and
RX\,J0852.0-4622, one finds a spatial correlation down to angular scales of
$\sim 0.1^\circ$, between the
X-ray emission and the TeV-band gamma-ray emission, with correlation factors
in excess of 70\%. This correlation suggests a common emission origin. The non-thermal 
X-ray emission is known to have structure on scales $\lesssim 0.01^\circ$, 
and it is the limited angular resolution and sensitivity of the current TeV observatories
that prevents a correlation analysis on the physically more relevant small scales.
Nevertheless, if the TeV gamma-ray
emission was of leptonic origin as suggested by the spatial
correlation, the spectra in X-rays and 
gamma rays should also be similar.
As hadronic gamma-ray production requires interaction
of the cosmic-ray nucleons with target nuclei, this emission will be
stronger for those SNRs located near or interacting with dense gas,
such as molecular clouds.  The TeV emission should be brightest in
those regions of the SNRs where the target density is highest.
 
In situ observations in the heliosphere show that collisionless shocks can accelerate particles.
The process of particle acceleration at SNR shocks is intrinsically efficient
\cite{kj05}. Thus, the shocks should be strongly modified,
because the energetic particles have a smaller adiabatic index and a much larger mean free path 
for scattering than does the quasi-thermal plasma. 
In addition, the particles at the highest 
energy escape, thus making energy losses significant and 
increasing the shock compression ratio \cite{be87}.
A fundamental consequence of particle acceleration at cosmic-ray modified shocks
is that the particle spectrum is no longer a power law, but a concave spectrum, 
as hard as $N(p)\propto p^{-1.5}$ at high momenta
\cite{ab06,vla06}. Gamma-ray observations in the GeV-TeV band appear to be the 
best means to measure the particle spectra and thus probe the
acceleration processes in detail.

Particle confinement near the shock is supported by self-generated magnetic 
turbulence ahead of and behind the shock that quasi-elastically scatters 
the energetic charged particles and thus makes their propagation diffusive. 
The amplitude of the turbulence 
determines the scattering frequency, and thus the acceleration 
rate \cite{drury83}. The instabilities by which cosmic rays drive turbulence 
in the upstream region were long thought to be weak enough so that
quasilinear approximations were realistic, i.e. $\delta B/B < 1$,
but recent research suggests that
the process by which streaming cosmic rays excite MHD
turbulence is different from that usually supposed, if the cosmic-ray 
acceleration is efficient.
The amplitude of the turbulent magnetic field may actually
exceed that of the homogeneous, large-scale field \cite{bl,lb}. More recent studies 
\cite{bell04} suggest
that ahead of the shock non-resonant, nearly purely growing modes of 
short wavelength may be more efficiently excited than resonant plasma waves.

The observation of narrow synchrotron X-ray filaments indicates 
that the magnetic field must be very strong at the particle
acceleration sites \cite{bamba,vink}, thus supporting the notion of magnetic-field 
amplification by cosmic rays.
Those strong magnetic fields will decay as the plasma convects
away from the forward shocks of SNRs, and it is an open question 
how far the regions of high magnetic field strength extend \cite{ply,cc07}.
The magnetic-field generation in shocks is also a 
candidate process for the creation of primordial magnetic fields in the cosmological context 
\cite{wid02,ss03,med06}. 
TeV-band gamma-ray observations, together with high-resolution X-ray studies, are the key to
understanding the generation of magnetic fields by energetic particles.  

\begin{figure*}[tb] \centerline{\psfig{file=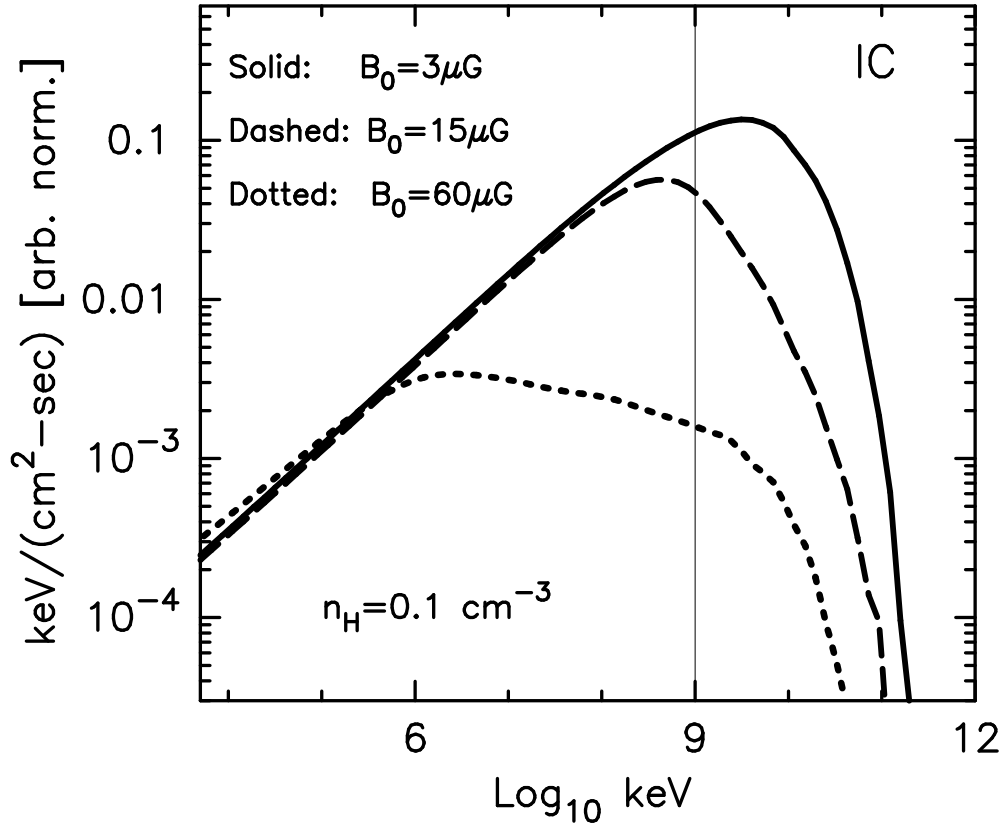,width=5.2in,clip=}} 
\caption{Expected GeV-TeV band gamma-ray emission from Inverse-Compton scattering
of the microwave background on highly relativistic electrons, according to recent
model calculations \cite{ell07}. Shown are three spectra for different values of the
magnetic field strength upstream of the SNR forward shock. For a high field strength
strong radiative losses and evolution make the IC spectrum significantly softer
above about 10~GeV, so it becomes similar to the expected gamma-ray spectrum
produced by energetic hadrons. The thin vertical line marks 1~TeV photon energy.} 
\label{ell-f12} \end{figure*}

If the acceleration efficiency is kept constant, a strong magnetic field would 
reduce the TeV-band gamma-ray emission arising
from IC scattering of energetic electrons relative to their synchrotron
X-ray emission, thus arguing against an IC origin of observable TeV-band emission.
Yet it would make the expected IC spectrum similar
to that of the hadronic pion-decay gamma rays \cite{voelk,ell07}, because strong energy
losses and evolution would produce a spectral change in the electron spectrum, 
as shown in Fig.~\ref{ell-f12}, and would also tie the spatial distribution of the 
gamma-ray emissivities even closer to that of the synchrotron X-rays. It is therefore mandatory
to combine sensitive spectral gamma-ray
measurements with a better angular resolution, so as to avoid confusion and
to effect discrimination between the hadronic and leptonic origin of the
gamma rays.

On the other hand, the pion-decay spectra in the GeV-TeV region, predicted by nonlinear particle 
acceleration models (e.g., \cite{bkv}), depend on 
uncertain parameters such as the ambient density and also somewhat on the strength of the 
interstellar magnetic field. RX~J1713-3946, the brightest shell-type SNR in the TeV band, 
harbors very little gas \cite{cassam}, thus making less likely a pion-decay origin of the
observed TeV-band emission. A robust discriminator, however, 
is the maximum photon energy. Since large magnetic fields produce severe radiation 
losses for electrons, there is a strong correlation between the ratio of maximum 
energy from ion-ion collisions to the maximum energy from IC and the magnetic field strength. 
The shape of the gamma-ray spectrum above $\sim$100~GeV also contains clues to the 
efficiency of the underlying acceleration process, and some SNRs, e.g., 
RX~J1713-3946 (see Fig.~\ref{ic443}), clearly show gamma-ray spectra too soft to be the result of 
efficient acceleration of cosmic-ray nucleons to the knee at 3~PeV \cite{huang}, where the spectrum
of Galactic cosmic rays starts to deviate from a simple power-law form. 

Since the massive stars of type O and B that explode as supernovae are 
predominantly formed in so-called OB associations, most SNRs \cite{hl80, sp90, hl06}
reside in superbubbles \cite{bru80, par04}, giant structures formed by the collective effect
of stellar winds and supernovae. Cosmic rays accelerated in superbubbles may achieve a higher particle
energy than those produced in isolated SNRs, possibly on account of stochastic acceleration processes
in the magnetic turbulence induced by the powerful multiple interacting supersonic stellar winds
\cite{cesarsky83,bykov01}. The winds from superbubbles, therefore, are a 
possible alternative cosmic-ray source class, and some aspects of the isotopic composition
of Galactic cosmic rays support their origin in superbubbles \cite{binns},
although the composition of the bulk of
the cosmic rays is that of the well-mixed interstellar medium \cite{meyer}, which somewhat limits
the role superbubbles can play as the main sources of Galactic cosmic rays. Even though a stellar cluster
may have already been seen in TeV-band gamma rays \cite{reimer}, it is very difficult to arrive at firm
theoretical estimates and interpretations for superbubbles because of their generally poorly known geometry and history, even though some of them are associated with
shell-like structures of atomic hydrogen \cite{mccg02}. There is the possibility of detecting the presence of high-energy heavy nuclei through their interaction with
the intense stellar radiation in clusters of massive stars. Nuclei with energies of a few PeV 
($10^{15}$~eV) will disintegrate upon collision with the starlight, and the subsequent 
de-excitation of the nuclear fragments gives rise to characteristic gamma-ray emission
with a distinct peak in power at about 10~TeV gamma-ray energy \cite{anch}.
 
Superbubbles in the Galaxy are typically 
several degrees in size (eg. the X-ray emitting Cygnus superbubble is \~15~$^{\circ}$ across), 
and therefore low surface brightness, confusion, and varying absorption
complicates their analysis in the radio, optical, and X-ray bands, so 
Galactic superbubbles may be incompletely cataloged \cite{tw05}. The Large Magellanic Cloud (LMC) 
is likely a better location to study superbubbles on account of its distance
(roughly 6 times the distance to the Galactic Center) and low foreground absorption. Numerous 
superbubbles have been found in the LMC \cite{dunne}, which are typically 10 arcminutes 
in apparent size, similar to Galactic SNRs. Nonthermal X-rays were observed from the outer
shell of the superbubble 30 Dor C \cite{bamba04} with a spectrum very similar to the 
nonthermal X-ray seen from Galactic SNRs like SN~1006, but with a luminosity 
about a factor of ten higher and an angular size of only a few minutes of arc, which 
partially compensates for the larger distance. The overall appearance
of superbubbles can, therefore, be likened to that of young SNRs, with one exception: 
the superbubbles are probably much older, so they can maintain efficient particle 
acceleration for $\sim 10^5$ years, in contrast to the $\sim 10^3$ years after which 
shell-type SNRs turn into the decelerating (Sedov) phase and gradually lose their ability
to efficiently accelerate particles. 

The TeV-band to keV-band nonthermal flux ratio of SNRs varies from object to object; 
that ratio is at least a factor of 20 lower for SN~1006 and Cas~A than for RX~J1713-3946. 
However, for superbubbles the flux ratio may actually be significantly higher than 
for isolated SNRs on account of the typical age of the objects, because 
the X-ray emitting electrons are severely loss-limited, whereas for gamma-ray-emitting ions 
that may not be the case. A relatively conservative TeV-band flux estimate can 
be made by taking the measured flux of nonthermal X-rays and
the flux ratio, as for RX~J1713-3946. With this, one would expect TeV-band gamma-ray emission 
from 30 Dor C at a level of a few milliCrab (1 Crab refers to the flux measured 
from the Crab Nebula: the standard candle in high-energy astrophysics), which is a factor 5-10 below the 
sensitivity threshold of present-generation imaging Cherenkov telescopes. Galactic
superbubbles may be much brighter, with about 1 Crab, but likely a few degrees in size, thus rendering their
detection and physical analysis equally difficult.

A future sensitive gamma-ray instrument is needed to
perform studies on a whole class of SNRs to 
finally understand the acceleration and interactions of both
energetic nucleons and electrons. It would also investigate in detail the new and exciting topic
of magnetic field amplification. Therefore, 
an advanced gamma-ray facility can, in conjunction
with current X-ray telescopes, provide detailed information on the
division of the energy budget in shocked SNR environs; namely, how
the global energetics is apportioned between cosmic ray electrons,
ions and magnetic field turbulence.  This is a principal goal that
will elucidate our understanding of plasma shocks, generation of
magnetic turbulence and cosmic ray acceleration in the cosmos.

\medskip
\subsubsection{Diffuse galactic emission}

\begin{figure*}[tb]
\centerline{\psfig{file=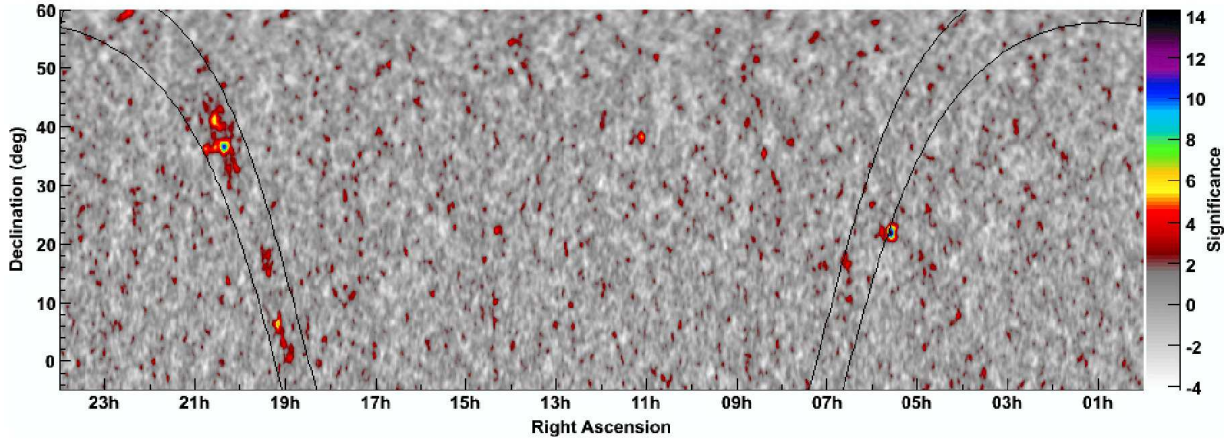,width=6.4in,clip=}} \caption{
A sky map from 5 years of Milagro data 
taking \cite{abdo}.  Clearly detected in this plot are the Crab Nebula and the 
Galactic ridge. The brightest portion of the inner Galaxy is the Cygnus region and we 
have strong evidence for an extended source embedded within the larger 
diffuse emission region.  The additional structure observed at lower 
Galactic latitudes has not yet been analyzed in detail and we cannot 
comment on the significance of any apparent features.} \label{milagro} \end{figure*}

In contrast to the case of SNRs, most of our knowledge of diffuse galactic gamma-ray emission
was obtained with survey-type instrumentation that combines a very large field-of-view with moderate
angular resolution of about one degree. Detectors like the Milagro instrument, that used the 
water-Cherenkov technique to measure gamma rays around 10~TeV energy, or the satellite experiment EGRET,
a pair-production instrument sensitive to GeV gamma rays that operated in the Nineties, 
fall into this category. Survey instruments
can provide good sensitivity to large-scale structures, but often suffer from confusion because the 
small-scale distribution of the signal cannot be determined, and so point sources and extended
emission cannot be reliably separated. On the other hand, atmospheric Cherenkov telescopes such as
H.E.S.S., MAGIC, and VERITAS offer a high angular resolution, so the angular structure of
compact sources can be properly determined; but they generally have a small field-of-view and a reduced
sensitivity for structures larger than a few degrees. The different characteristics of survey
instruments and high-resolution cameras are evident in the scientific results of existing experiments.

EGRET has produced an all-sky map of the gamma-ray sky up to 10 GeV; at these
energies inverse Compton (IC) scattering is still a major component of diffuse emission,
possibly even dominant. NASA's next-generation experiment, Fermi, will
clarify the nature of excess emission seen with EGRET 
at a few GeV, dubbed the GeV excess \cite{hunter}, 
produce an allsky map of GeV-scale diffuse gamma-ray emission, 
and also extend the coverage to 100 GeV. While at 1~GeV the statistical accuracy will be very high, with more 
than a hundred detected gamma rays per year and angular resolution element, the angular resolution
as measured through angle around the true photon direction for 68\% containment still exceeds 0.5~degrees,
so confusion will be an issue.
At higher energies, around 30~GeV, the angular resolution is better than $0.1^\circ$, but we can expect only
about one detected photon per $0.1^\circ$ resolution element through the 5-year mission, so
the angular resolution cannot be fully exploited. Fermi will provide
invaluable spectral information on the diffuse Galactic gamma-ray emission in the GeV band with 
degree-scale angular resolution, but TeV-band measurements will produce complementary images and spectra 
with very high angular resolution for selected regions of the sky that will be particularly useful 
where imaging with Fermi suffers from confusion. GeV- and TeV-band observations can be combined to extract
the information required to understand the propagation of energetic particles in the Galaxy.

The IC contribution to the diffuse Galactic gamma-ray emission
can be large and not easy to separate from that of pion-decay. The
separation of the diffuse gamma-ray signal into the contributions of cosmic-ray ions and those of
electrons is desirable, because the propagation properties of the two particle populations is different.
Also, measurements of the isotopic composition of cosmic rays near earth with
appropriate particle detectors such as, e.g., PAMELA \cite{malvezzi} 
allows us to additionally constrain the propagation history of cosmic-ray ions, although it appears
very difficult to both fit the EGRET data and the locally measured spectra of cosmic-ray ions
and electrons \cite{andy04}.

For gamma rays with energy above 10 TeV, the electron energies have to be at least a few 
tens of TeV, but in view of the rapid energy losses it is probable that the electrons do
not have the time to propagate away from their
acceleration sites; hence, IC is of much less importance for the diffuse emission at those very high energies.
An all-sky map above 10 TeV would provide a 'clean' view of the distribution 
and spectrum of cosmic-ray hadrons over the whole Galaxy. Such a skymap could 
provide the key to the origin of cosmic-ray hadrons, in particular when it could be combined with 
information on the intensity of neutrinos. 

A measurement of the intensity of diffuse emission at TeV energies would be extremely valuable, 
provided one is able to separate truly diffuse
emission from individual sources such as pulsar-wind nebulae. To date, 
the Milagro collaboration
reports evidence of TeV-scale gamma-ray emission from the Galactic plane, in particular the 
Cygnus region (see Figure~\ref{milagro}). 
The intensity measured with Milagro at 12~TeV is 70~Crab/sr (about 0.02 Crab/deg$^2$)
and thus extremely sensitive to the point source content. For comparison, the intensity of 
the 14 new sources detected during the H.E.S.S. survey of the inner Galaxy
\cite{hess-survey}, if they were unresolved, would
be 17~Crab/sr above 200~GeV. Assuming the measured spectrum extends to 12~TeV, the equivalent
intensity of the 14 H.E.S.S. sources at 12~TeV, which is more relevant for a comparison 
with the Milagro result, would be 140~Crab/sr, i.e. twice the intensity observed with Milagro. 
The source density in the region observed with Milagro is probably smaller, but there are also sources
not seen with H.E.S.S. or known prior to the survey, and therefore
a significant fraction of the Milagro result will be due to unresolved sources, and confusion is
a substantial problem.

\begin{figure*}[tb]
\centerline{\psfig{file=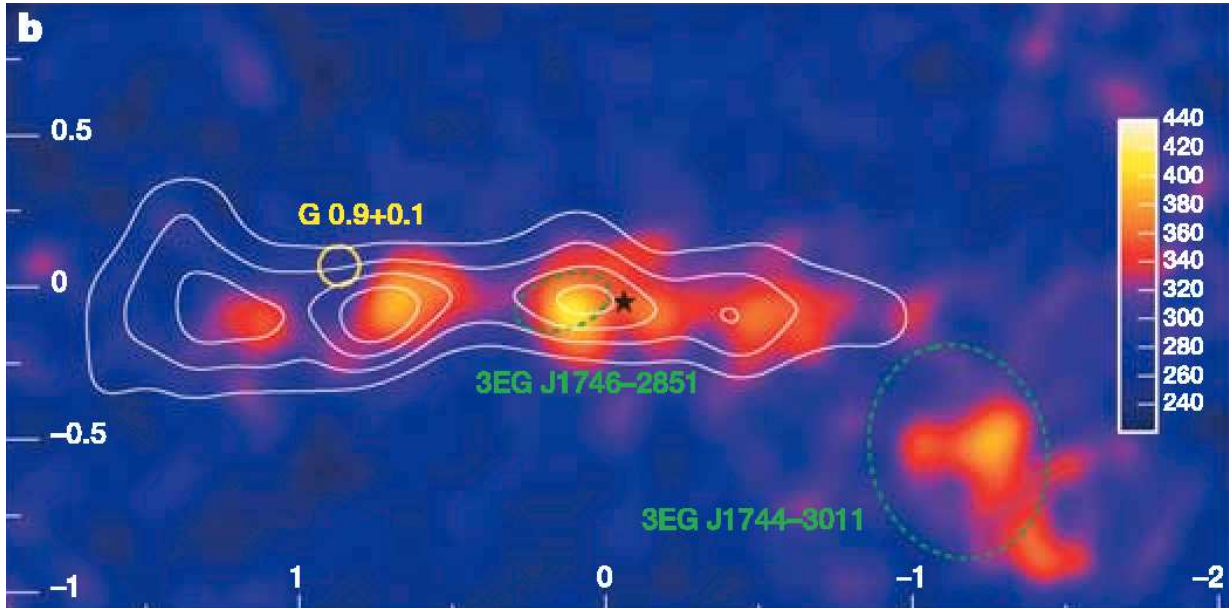,width=6.4in,clip=}} \caption{
H.E.S.S. gamma-ray count map after subtraction of two
bright point sources. The white contour lines indicate the column density of 
molecular gas traced by CS line emission.} \label{hess} \end{figure*}

The H.E.S.S. collaboration has published a map and the spectrum of diffuse emission from the 
inner degree of the Galaxy, after subtracting two dominating point sources (see Figure~\ref{hess}). 
In a fit of the observed spectrum as $dN/dE\propto E^{-\gamma}$, the spectral index is 
$\gamma=2.3\pm 0.08$, much harder than expected and measured anywhere else, as cosmic ray
ions with a spectrum as directly measured at earth should produce a gamma-ray spectrum with 
$\gamma\simeq 2.7$. The measured intensity 
corresponds to 590 Crab/sr at a TeV and is about the highest one may expect anywhere 
in the Galaxy based on the intensity distribution of GeV-band gamma rays.

\subsection{What measurements are needed?}

%\medskip
\subsubsection{Supernova remnants}

With multi-waveband data, it
is possible to provide quantitative constraints on the particle acceleration
mechanism.  Because the maximum IC power output from these objects is
expected to be in the TeV region, TeV observations provide information
unavailable via any other means.  High-resolution maps and accurate
spectra of the TeV emission, when compared with data from other wavebands,
will permit estimation of the magnetic field and the maximum energy of
the accelerated particles.  Comparison of the maps from various
wavelengths will increase our understanding of the diffusion and
lifetimes of the highly energetic electrons. 

We stress the importance of GeV-band
gamma-ray data that will be shortly provided by Fermi. However, the sensitivity of
Fermi at 10--100~GeV is limited: if we extrapolate the TeV-band spectrum of RX~J1713-3946 
(see Fig.~\ref{ic443}) to lower energies as $dN/dE\propto E^{-2}$, or a flat line in the figure, then
we expect Fermi to detect two photons per year with energy above 100~GeV from the entire SNR. 
Above 10~GeV, Fermi would find 20 photons per year, so even after five years, a Fermi gamma-ray excess map
would have much lower statistical accuracy than the existing H.E.S.S. map; the single-photon resolution
is also worse. At energies below 10~GeV, the number of Fermi-detected photons increases, 
but the angular resolution deteriorates. Fermi will perform important
studies of shell-type SNRs, but TeV-band measurements will provide complementary and, in many cases, 
richer images and spectra.
 
A key for any future VHE observatory will be to unambiguously
disentangle the emission from electronic versus hadronic cosmic rays.
Spectral studies may help arriving at a discrimination between 
gamma rays from electrons and those produced by hadrons, but they are not sufficient. 
TeV gamma rays from IC scattering of the microwave background should have a spectral shape
that reflects that of synchrotron X-rays below approximately 1 keV, where 
the discrimination of synchrotron emission and thermal radiation of ordinary hot gas
is often difficult and requires a very good angular resolution.

\begin{figure*}[tb]
\centerline{\psfig{file=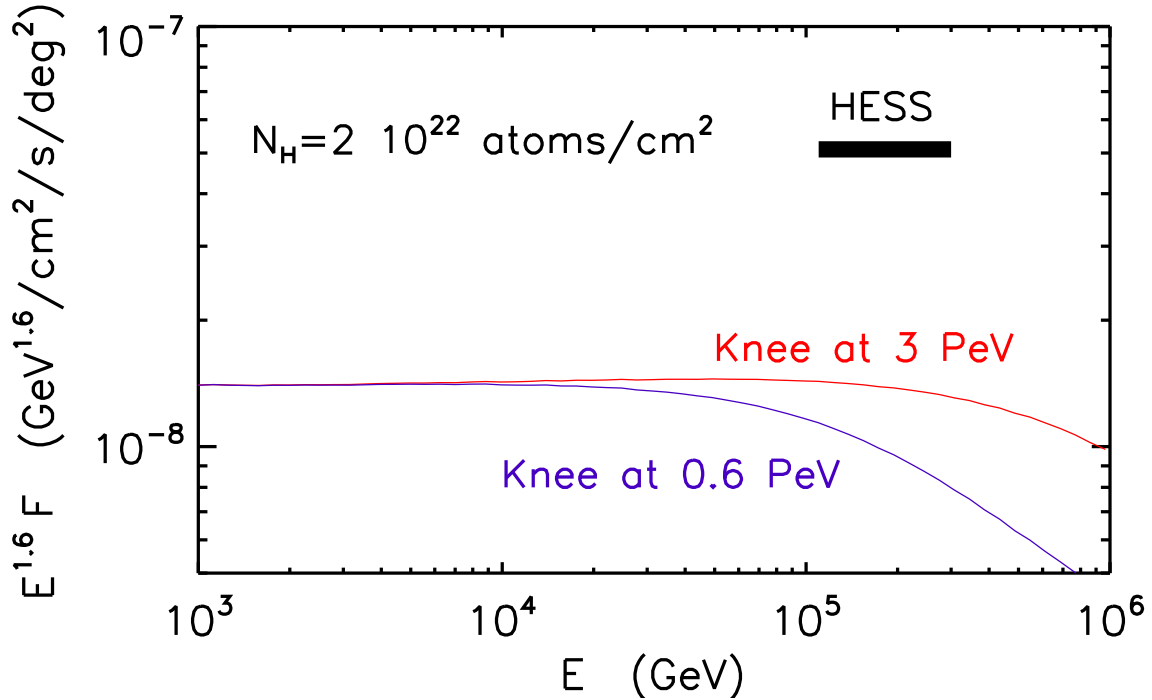,width=6.0in,clip=}} \caption{
Shown as red line is the intensity of diffuse Galactic gamma rays, multiplied with $E^{1.6}$, 
for a standard cosmic-ray spectrum with the knee at 3~PeV and one of the Orion molecular clouds. 
If near some molecular gas complex the
knee was at 0.6~PeV, the spectrum of diffuse gamma rays from that region would follow the blue line. 
Observing a location dependence of the knee energy would provide important clues on the nature of the knee,
as do similar measurement for individual sources of cosmic rays (e.g. \cite{huang}).
The black bar indicates 
an estimate of the current H.E.S.S. sensitivity in the 100--300~TeV band, based on published
spectra of RX~J1713-3946. An increase by a factor 10 in sensitivity around 200~TeV would be needed to
discriminate the blue and the red curve.} \label{knee} \end{figure*}

On the other hand, TeV gamma rays of hadronic origin reflect the spectrum of
energetic nuclei at about 1--100~TeV energy. If the SNR in question 
accelerates hadronic cosmic rays to energies beyond the knee at 3~PeV, then we should see 
a continuation of the gamma-ray emission up to 100 TeV and beyond, which would be a good
indication of a hadronic origin of gamma rays (see also Fig.~\ref{knee}). 
It will therefore be important to maintain
a sensitivity up to and beyond 100~TeV. We also note an obvious relation to neutrino astrophysics:
all gamma-ray sources at the Crab flux level that do not cut off below 100~TeV energy should be observable
with neutrino detectors \cite{kis+bea}, if the gamma-ray emission arises from interactions of 
energetic nuclei.
In the near term, only IceCube at the South Pole will be large enough
to observe Galactic neutrino sources such as SNRs.  Since it must look through
Earth, it is very important that it be paired with sensitive gamma-ray
instruments in the Northern hemisphere.  The H.E.S.S. experiment and its possible successor, CTA, 
planned for the same site in Namibia, can observe many Southern gamma-ray sources,
but they won't be paired with an adequately sensitive km$^3$-scale Mediterranean neutrino detector until
much later. The US-led VERITAS experiment currently observes sources in the Northern sky.
A more sensitive successor could optimally exploit the scientific opportunities that lie in
the synergies with IceCube. An additional future survey-type instrument could search 
for very extended sources and serve as a pathfinder for high-angular-resolution observations with
the atmospheric Cherenkov telescopes.
 
High-resolution imaging of the TeV emission, combined with good spectral information
is, therefore, required.
TeV emission from hadronic interactions should
trace the distribution of target material, while
TeV emission from electrons should be well correlated with non-thermal
X-ray emission (see Figs.~\ref{ic443}). 
Because a significant fraction of the non-thermal X-ray intensity is organized in thin
filaments, arcminute-scale resolution in the TeV band combined with the
appropriate sensitivity would permit a clear separation of hadronic and leptonic emission,
and would allow a direct measurement of the magnetic field strength at the forward shock of 
SNR, and hence a clean assessment of the efficacy of magnetic-field amplification by energetic 
particles. The required angular resolution is, therefore, a factor 3--5 better than 
what is currently achieved with H.E.S.S. and VERITAS. The sensitivity needed to derive 
well-defined spectra with angular resolution below $0.1^\circ$ is about a factor of 10 higher 
than that afforded by the current generation of atmospheric \v Cherenkov telescopes. This
is also the sensitivity likely needed to detect TeV-band gamma rays from superbubbles. 

Non-detection of hadronically produced gamma rays
would require either a very steep source spectrum, inconsistent with
that needed to produce the local spectrum, or a greatly reduced
cosmic-ray intensity, inconsistent with the energy budget for cosmic
rays.  Either of these possibilities would lead to
serious revisions in our understanding of the origin of cosmic rays. 
Detection of TeV photons from hadronic cosmic rays would immediately
constrain the spectrum and total energy budget of the cosmic rays, and
would provide invaluable constraints on the relative acceleration efficiency
of electrons and protons or other ions in shocks. This may help resolve  
the hundred-year-old question of the origin of cosmic rays, and will yield important 
information on shock physics that can be used in other shock systems. If hadronic
cosmic rays are accelerated in shocks produced by the winds from OB
associations, the TeV photons produced by those cosmic rays should,
again, trace the distribution of target material. The angular
resolution requirements are similar to those discussed for supernova
remnants.

\subsubsection{Diffuse galactic emission}

Currently most pressing questions are the following:
Are cosmic rays above the knee at 3~PeV, where the spectrum of local cosmic rays
considerably steepens, really Galactic in origin? What is the origin of the
knee? Is 
the knee a source property, in which case we should see a corresponding
spectral feature in the gamma-ray spectra of cosmic-ray sources, or the result of propagation, so we
should observe a knee that is potentially dependent on location, because the propagation properties
depend on position in the Galaxy? 

Another series of questions concerns cosmic-ray electrons, whose source power is significant, but whose
spectrum above 1 TeV is essentially unknown. What is the distribution of cosmic-ray electrons at 
energies beyond 1 TeV? Measuring electron spectra inside and outside their sources
carries direct information on the particle acceleration rate, and thus on the nature of the 
acceleration process, as well as on the propagation properties of cosmic rays up to the knee.

Via their gamma-ray emission, we would therefore wish to independently measure

\begin{itemize}
\item the spectrum and flux of cosmic nuclei, 
which are expected to produce a gamma-ray signal that
largely correlates with the density of interstellar gas. The expected intensity on a half-degree scale
in, e.g., the Cygnus region is 60 Crab/sr (0.02 Crab/deg$^2$) in the 100~GeV--1~TeV band, and
40 Crab/sr (0.013 Crab/deg$^2$) above 1~TeV. 
Note that gamma rays with energy higher than about 100 TeV map 
the cosmic-ray nuclei spectrum around the knee, so an increase by at least
a factor 10 in sensitivity and a
good energy resolution up to and beyond
100~TeV is required to potentially prove a location dependence of the knee (see also Fig.~\ref{knee}).
At these energies, pair production with ambient radiation can attenuate the gamma-ray signals as they
travel across the Galaxy, but observations of relatively nearby complexes of molecular clouds
would ensure that absorption in the Galaxy is 
negligible and that the intensity measurement can be made by integration over typically a 
square-degree in solid angle.
One should note that a high angular resolution is nevertheless needed 
for those measurements, both to account for point-source contributions and to verify the spatial correlation
of the signal with the distribution of atomic
molecular gas, which is known on scales $\lesssim 0.1^\circ$. 

\item the spectrum and flux of cosmic electrons, which will produce a patchy and spectrally variable
gamma-ray signal that does not correlate with the gas density but may have structure on a 
1--3~$^{\circ}$ scale. The intensity is impossible to estimate without insight into the nature of 
the EGRET GeV excess, but may be stronger than the hadronic emission in the 100~GeV--1~TeV band.

\item the point-source content of the gamma-ray signal to properly separate sources from truly 
diffuse emission.
\end{itemize}

None of these measurements requires a very low energy threshold, though one would wish to 
not have a large gap to the energy range accessible with Fermi, which will make reliable 
measurements up to about 50~GeV gamma-ray energy. The measurement of hadronic and, in particular, 
leptonic gamma rays chiefly requires advances in both the effective area and, in particular,
the background rejection of future observatories. The field-of-view, in which sensitive 
gamma-ray observations are taken, is of minor importance,
as long at it is at least 4~$^{\circ}$ in diameter. This requirement arises
from the necessity to independently determine a reliable gamma-ray zero level, which is best done
by having the gamma-ray-sensitive field-of-view
cover both the source, e.g. an SNR or a molecular cloud complex, and empty regions surrounding it 
for background measurements. 
%One should note that in atmospheric 
%\v Cerenkov telescopes
%the gamma-ray sensitive field-of-view is typically about 2~$^{\circ}$ narrower than
%the optical field-of-view of the telescope, somewhat depending on the gamma-ray energy.

As a byproduct, one would also be interested in measuring the direct \v Cerenkov light of local cosmic rays, 
which provides unique information on the cosmic-ray composition in the PeV energy range. 
Although cosmic rays form the primary background for ground-based gamma-ray 
detectors, this background can be used to make a 
unique measurement of cosmic ray composition.  Recently, the H.E.S.S. 
collaboration has measured the 
direct \v Cherenkov light of local cosmic rays \cite{hess-direct}, which provides unique 
information on the cosmic-ray composition in the PeV energy range \cite{kieda2001}. 
This method is more direct than that 
used in extensive air shower experiments because it avoids the 
dependence on hadronic simulations in identifying the primary particle 
type. Also, the main air showers display strong 
statistical fluctuations in their evolution, thus making the observation of 
direct \v Cherenkov light a much more precise measurement. 
The current atmospheric \v Cherenkov arrays like H.E.S.S. or VERITAS lack the 
angular and timing resolution to fully exploit this method, and
a future instrument with 0.01-degree image pixel 
resolution and nanosecond time resolution 
could further improve the determination of the cosmic-ray composition at 
high energies. 

\subsection{What is the required instrument
performance?}

For the study of SNR 
the two key instrument parameters are angular resolution and a
sufficiently high count rate to effectively exploit the angular resolution. 
The
required angular resolution can be estimated from the known angular sizes of the
non-thermal X-ray filaments and of the dense molecular clouds and shock
regions.  For the closer SNRs, the typical angular resolution required
is about one arcminute. A key point is that a sufficient number of gamma rays 
must be detected to make effective use of the angular resolution.  The
detection rate can be increased relative to current instruments either
by increasing the effective area or by reducing the energy threshold.
The goal should be to image several SNRs with arcminute-level resolution with 
a minimum of 150 events in each image bin,
so reliable spectra can be reconstructed.

To maximize the scientific return for Galactic sources, a future
instrument should be located at sufficiently southern latitude to give good
coverage of a large fraction of the Galactic plane extending to the
inner Galaxy. At the same time it is desirable that a large overlap is maintained with the
coverage of neutrino experiments such as IceCube, which makes a Southern location less advantageous. 

To achieve scientifically significant observations of the diffuse Galactic gamma-ray emission with
the next-generation instrument, the angular resolution is important, but does not need to be as good as for
observations of SNR. Mainly one needs to model and subtract individual sources. A good
angular resolution is also needed to find intensity that correlates with the gas distribution. 

It is mandatory to achieve a very high sensitivity for extended emission.
Given that a number of emission components have to be fitted in parallel
and that, at least for the diffuse emission, the data are background-dominated, a
strongly improved background rejection is required to achieve the desired sensitivity. 
A large aperture alone appears insufficient, as it is necessary to achieve both large event 
numbers and a very low background contamination level.

The instrument requirements can thus be summarized as follows:

\begin{itemize}
\item an angular resolution of $\le 0.02^\circ$ at 1 TeV.
\item for the bright parts of SNRs at least 150 gamma rays in each image bin for a 
reasonable observing time.
\item a sensitivity for extended emission that is significantly better than 10 Crab/sr above a 
TeV and better than 15 Crab/sr below a TeV.
\item maintain a high sensitivity up to and possibly beyond 100~TeV.
\item a good energy resolution of $\delta E/E\lesssim 15\%$ at all energies.
\item a gamma-ray sensitive field-of-view of at least 4~$^{\circ}$ in diameter. Bigger is better, 
but tens of degrees are not needed.
\end{itemize}

\end{document}